\documentclass[a4paper,11pt]{article}

\usepackage{authblk}
\usepackage{algorithm}
\usepackage{algorithmic}
\usepackage{multirow}
\usepackage{graphicx}
\usepackage{booktabs}
\usepackage{balance}
\usepackage{epsfig}
\usepackage{epstopdf}
\usepackage{amsmath}
\usepackage{amsthm}
\usepackage{amsfonts}
\usepackage{subcaption}
\usepackage{wasysym}
\usepackage{microtype}
\usepackage{rotating}
\usepackage{tikz}
\usepackage{txfonts}
\usepackage{fancyhdr}
\usepackage{url}
\usepackage{hyperref}
\usetikzlibrary{positioning}
\usetikzlibrary{decorations.pathreplacing}
\tikzset{
	mybrace/.style={decorate,decoration={brace,aspect=#1}}
}

\newcommand{\N}{\mathbb{N}}

\newcommand{\F}{\mathbb{F}}

\newtheorem{definition}{Definition}
\newtheorem{lemma}{Lemma}

\providecommand{\keywords}[1]{\textbf{\textit{Keywords }} #1}

\begin{document}

\title{Exploring Semi-bent Boolean Functions Arising from Cellular Automata}

\author[1]{Luca Mariot}
\author[2]{Martina Saletta}
\author[2]{Alberto Leporati}
\author[3]{Luca Manzoni}

\affil[1]{{\normalsize Cyber Security Research Group, Delft University of Technology,
              Mekelweg 2, Delft, The Netherlands} \\
  
  {\small \texttt{l.mariot@tudelft.nl}}}

\affil[2]{{\normalsize DISCo, Universit\`{a} degli Studi di Milano-Bicocca,
    Viale Sarca 336/14, 20126 Milano, Italy} \\
  
  {\small \texttt{alberto.leporati@unimib.it, m.saletta1@campus.unimib.it}}}

\affil[3]{{\normalsize Dipartimento di Matematica e Geoscienze, Universit\`{a} degli Studi di Trieste,
    Via Valerio 12/1, 34127 Trieste, Italy} \\
  
  {\small \texttt{lmanzoni@units.it}}}

\maketitle

\begin{abstract}
Semi-bent Boolean functions are interesting from a cryptographic standpoint, since they possess several desirable properties such as having a low and flat Walsh spectrum, which is useful to resist linear cryptanalysis. In this paper, we consider the search of semi-bent functions through a construction based on cellular automata (CA). In particular, the construction defines a Boolean function by computing the XOR of all output cells in the CA. Since the resulting Boolean functions have the same algebraic degree of the CA local rule, we devise a combinatorial algorithm to enumerate all quadratic Boolean functions. We then apply this algorithm to exhaustively explore the space of quadratic rules of up to $6$ variables, selecting only those for which our CA-based construction always yields semi-bent functions of up to $20$ variables. Finally, we filter the obtained rules with respect to their balancedness, and remark that the semi-bent functions generated through our construction by the remaining rules have a constant number of linear structures.
\end{abstract}


\section{Introduction}
\label{sec:intro}
\emph{Cellular Automata} (CA) represent an appealing approach to the design of cryptographic primitives. Indeed, starting from the 80s, CA have been extensively investigated for designing \emph{Pseudo-Random Number Generators} (PRNGs)~\cite{wolfram85,seredynski04,manzoni18}, \emph{S-boxes}~\cite{szaban08,ghoshal18,mariot19} and \emph{secret sharing schemes}~\cite{delrey05,mariot14,mariot18}, among other things.

In this work, we consider the use of CA for the construction of \emph{Boolean functions} with interesting cryptographic properties. Boolean functions are cryptographic primitives that play an important role in the design of \emph{stream ciphers}, where they may be used to combine or filter the output of linear feedback shift registers (LFSR) to construct a keystream, and in \emph{block ciphers}, where they constitute the coordinates of S-boxes. Previous research~\cite{leporati14,formenti14} focused on the investigation of CA local rules as Boolean functions, selecting those with the best cryptographic properties to withstand particular attacks when used in a CA-based PRNG. In this work we adopt a different viewpoint, which spawns from the following question: given a Boolean function of $m$ variables with good cryptographic properties, is it possible to derive new functions from it with a larger number of variables and analogous properties by using a CA?

More specifically, the construction that we investigate in this paper employs an initial $m$-variable Boolean function as the local rule of a CA of $n \ge m$ cells. Then, a new function of $n$ variables is constructed by applying the CA global rule and by computing the XOR of the CA cells in the output configuration. In this way, one can generate an infinite family of Boolean functions starting from the initial local rule by simply adding more cells to the CA. Techniques for generating new Boolean functions from existing ones are also called \emph{secondary} (or \emph{recursive}) \emph{constructions}, and only few of them are known in the related literature, none of which are based on CA (see e.g.~\cite{carlet10} for a survey). 
Our analysis focuses on the particular case of \emph{semi-bent} Boolean functions, which have interesting cryptographic properties such as high nonlinearity. In particular, we are interested in finding semi-bent functions which generate larger semi-bent functions when plugged as local rules in our CA-based construction. As a first basic result, we show that our construction preserves the algebraic degree of the local rule. We thus design a combinatorial algorithm based on the \emph{Algebraic Normal Form} representation to enumerate all Boolean functions of a fixed degree. For our experiments, we use our algorithm to enumerate all \emph{quadratic} functions of $3\le m \le 6$ variables, and among them we select only those that generate semi-bent functions of up to $n=20$ variables through our CA construction. The first remarkable finding is that for $m=4$ variables our construction always fails, i.e. no quadratic rule of $4$ variables is able to generate semi-bent functions of up to $n=20$ variables. By focusing on the balanced rules of $3$, $5$ and $6$ variables over which the construction works, we finally remark that they all have a constant number of non-trivial \emph{linear structures}, namely $1$ when the number of variables is odd, and $3$ when it is even.

The rest of this paper is organized as follows. Section~\ref{sec:background} covers the basic definitions concerning Boolean functions and their cryptographic properties. Section~\ref{sec:construction} introduces the CA model considered in this work and defines our CA-based construction of Boolean functions, while Section~\ref{sec:algorithm} describes the search algorithm used to enumerate functions of a fixed degree. Section~\ref{sec:results} presents the results of our exhaustive search experiments on the spaces of quadratic local rules. Finally, Section~\ref{sec:conclusions} concludes the paper and points out some open problems concerning our construction for future research.

\section{Background on Boolean Functions}
\label{sec:background}
In what follows, let $\F_2 = \{0,1\}$ denote the finite field of two elements and let  $\F_2^n$ be the $n$-dimensional vector space over $\F_2$. The \emph{support} of $x \in \F_2^n$ is defined as $supp(x) = \{i : x_i \neq 0 \}$, while the \emph{Hamming weight} of $x$ is $w_H(x)=|supp(x)|$, i.e. the number of $1$s in $x$.

A \emph{Boolean function} of $n \in \N$ variables is a mapping $f: \F_2^n \to \F_2$, with its \emph{truth table} being the $2^n$-bit string $\Omega_f$ that specifies the output value of $f$ for each of the vectors in $\F_2^n$, in lexicographic order. A function $f$ is called \emph{balanced} if its truth table is composed of an equal number of 0s and 1s, i.e. if $w_H(\Omega_f) = 2^{n-1}$. Balancedness is a fundamental cryptographic property that Boolean functions used in stream and block ciphers should satisfy to resist statistical attacks.

Besides the truth table, a second unique representation of a Boolean function $f:\F_2^n\to \F_2$ commonly used in cryptography is the \emph{Algebraic Normal Form} (ANF), which is defined as the following multivariate polynomial over the quotient ring $\mathbb{F}_2[x_1,\cdots,x_n]/(x_1^2 \oplus x_1, \cdots, x_n^2 \oplus x_n)$:
\begin{equation}
    P_f(x) = \bigoplus_{I \in 2^{[n]}} a_I \left( \prod_{i \in I} x_i \right) \enspace ,
\end{equation}
where $2^{[n]}$ is the power set of $[n] = \{1,\cdots,n\}$. The \emph{algebraic degree} of $f$ is the cardinality of the largest subset $I \in 2^{[n]}$ in its ANF such that $a_I \ne 0$. In particular, \emph{affine functions} are defined as those Boolean functions with degree at most $1$. As a cryptographic criterion, the algebraic degree should be as high as possible. The vector of the ANF coefficients $a_I$ and the truth table of $f$ are related by the \emph{M{\"o}bius transform}:
\begin{equation}
  \label{eq:mobius}
  f(x) = \bigoplus_{I \in 2^{[n]}: I \subseteq supp(x) } a_I \enspace ,
\end{equation}

Another representation used to characterize several cryptographic properties of Boolean functions is the Walsh transform. Formally, the \emph{Walsh transform} of a Boolean function $f: \F_2^n \to \F_2$ is defined for all $a \in \F_2^n$ as:
\begin{equation}
    \label{eq:walsh}
    W_f(a) = \sum_{x \in \F_2^n} (-1)^{f(x) \oplus a\cdot x} \enspace ,
\end{equation}
where $a\cdot x = \bigoplus_{i=1}^n a_ix_i$ is the \emph{scalar product} of the vectors $a$ and $x$. A function $f$ is balanced if and only if the Walsh coefficient over the null vector is zero, i.e. $W_f(0) = 0$. More in general, the coefficient $W_f(a)$ measures the \emph{correlation} between $f$ and the linear function $a\cdot x$. Thus, the Walsh transform can be used to compute the \emph{nonlinearity} of a Boolean function $f$, which is defined as the minimum Hamming distance of $f$ from the set of all affine functions. In particular, the nonlinearity of $f$ equals
\begin{equation}
    \label{eq:nl}
N_f = 2^{n-1} - \frac{1}{2}\cdot \max_{a \in \F_2^n}\{|W_f(a)|\} \enspace .
\end{equation}

For cryptographic applications, the nonlinearity of the involved Boolean functions should be as high as possible. From Equation~\eqref{eq:nl}, this means that the maximum absolute value of the Walsh transform should be as low as possible. By \emph{Parseval relation}, this can happen only when all Walsh coefficients have the same absolute value $2^{\frac{n}{2}}$, yielding the \emph{covering radius bound}: $N_f \le 2^{n-1} - 2^{\frac{n}{2} - 1}$. Functions satisfying this bound are called \emph{bent}, and they exist only when $n$ is even. Unfortunately such functions are not balanced, since $W_f(0) = \pm 2^{\frac{n}{2}}$, and thus they cannot be used directly in the design of stream and block ciphers. For $n$ odd, the \emph{quadratic bound} is given by $N_f \le 2^{n-1} - 2^{\frac{n+1}{2} - 1}$, and it can be always achieved by functions of algebraic degree $2$.

\emph{Plateaued functions} represent an interesting generalization of bent functions, since they can also be balanced while still retaining high nonlinearity. Formally, a Boolean function $f: \F_2^n \to \F_2$ is \emph{plateaued} if its Walsh transform takes only three values, i.e. if $W_f(a) \in \{-\lambda, 0, +\lambda\}$ for all $a \in \F_2^n$. In particular, a plateaued function is \emph{semi-bent} if $\lambda = 2^{\frac{n+1}{2}}$ for $n$ odd and $\lambda = 2^{\frac{n+2}{2}}$ for $n$ even.  This means that the nonlinearity of a semi-bent function equals $2^{n-1} - 2^{\frac{n-1}{2}}$ when $n$ is odd and $2^{n-1} - 2^{\frac{n}{2}}$ when $n$ is even. Hence, semi-bent functions reach the quadratic bound for nonlinearity when $n$ is odd.

We conclude this section by recalling the concept of \emph{linear structures}. Given a Boolean function $f: \F_2^n \to \F_2$, the \emph{derivative} of $f$ with respect to $b \in \F_2^n$ is defined as $D_bf(x) = f(x) \oplus f(x \oplus b)$. Then, $b$ is called a \emph{linear structure} for $f$ if the derivative is a constant function, that is, if $D_bf(x) = 0$ for all $x \in \F_2^n$ or $D_bf(x) = 1$ for all $x \in \F_2^n$. Remark that the null vector is a trivial linear structure, since $D_0f(x) = f(x) \oplus f(x \oplus 0) = 0$ for any Boolean function $f$. Ideally, the number of linear structures in Boolean functions used for stream and block ciphers should be as low as possible.

\section{Our Construction}
\label{sec:construction}
We start by introducing the CA model considered in this work.
\begin{definition}
    \label{def:ca}
    Let $f:\F_2^m \to \F_2$ be a Boolean function of $m$ variables, and $n\ge m$. A \emph{Cellular Automaton} (CA) of $n$ cells and \emph{local rule} $f$ is a vectorial function $F: \F_2^{n} \to \F_2^{n-m+1}$ defined for all $x \in \F_2^n$ as:
            \[
                F(x_1, x_2, \cdots, x_n) = (f(x_1, \cdots, x_m), \cdots, f(x_{n-m+1}, \cdots, x_n)) \enspace .
            \]
\end{definition}
A CA can thus be seen as a vectorial Boolean function where each coordinate function $f_i:\F_2^m \to \F_2$ corresponds to the local rule $f$ applied to the \emph{neighborhood} $(x_i,\cdots,x_{i+m-1})$. This rule is applied just up to the coordinate $n-m+1$, meaning that the size of the input array shrinks by $m-1$ cells. We remark that Definition~\ref{def:ca} corresponds to the \emph{No Boundary CA} model studied in~\cite{mariot19} for CA-based S-boxes, and in~\cite{mariot20} for mutually orthogonal Latin squares. Since the local rule $f: \F_2^{m} \rightarrow \F_2$ is a Boolean function, it can be defined by a truth table $\Omega_f$ of $2^{m}$ bits. In the CA literature, the truth table of a local rule is usually represented by its \emph{Wolfram code}, which amounts to the decimal value of $\Omega_f$ seen as a binary number.

\begin{figure}[t]
    \centering
    \includegraphics[scale=0.7]{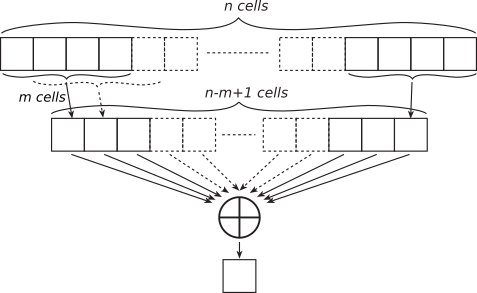}
    \caption{Representation of our CA-based construction for Boolean functions.}
    \label{fig:constr}
\end{figure}

We can now define our construction of Boolean functions based on the no-boundary CA model discussed above.
\begin{definition}
\label{def:constr}
Let $F: \F_2^n \to \F_2^{n-m+1}$ be a CA of length $n\ge m$ equipped with the local rule $f:\F_2^m \to \F_2$. Then, the Boolean function induced by $f$ through the CA $F$ is the $n$-variable function $f^*: \F_2^n \to \F_2$ defined for all $x \in \F_2^n$ as:
\begin{equation}
  \label{eq:constr}
  f^*(x) = \bigoplus_{i=1}^{n-m+1} f(x_i,\cdots,x_{i+m-1}) = f(x_1,\cdots,x_m) \oplus \cdots \oplus f(x_{n-m+1},\cdots,x_n) \enspace .
\end{equation}
\end{definition}
In other words, the construction consists in first applying the CA vectorial function $F$ induced by the local rule $f$ to the input vector $x \in \F_2^n$; then, the value of the constructed function $f^*$ is obtained by computing the XOR of all the output cells of the CA. Figure~\ref{fig:constr} gives a schematic depiction of how the construction works.
Using the terminology of the Boolean functions literature~\cite{carlet10}, the construction of Definition~\ref{def:constr} may be classified as a \emph{secondary construction}, since it starts from a known function $f$ of $m$ variables used as a CA local rule, and generates a new function $f^*$ of $n$ variables from it. In particular, our construction gives rise to an \emph{infinite family} of Boolean functions, since $f^*$ can be defined for any number of variables $n\ge m$ by simply adding $n$ cells to the CA.

Secondary constructions are mainly employed to generate new Boolean functions from old ones with analogous cryptographic properties. For example, Rothaus's construction~\cite{rothaus76} starts from three bent functions of $n$ variables, whose sum is also bent, and produces a new bent function of $n+2$ variables. We thus need to analyze which properties are preserved by our construction. The next lemma shows that the algebraic degree is one such property:
\begin{lemma}
\label{lm:alg-deg}
Let $f: \F_2^m \to \F_2$ be a Boolean function of $m$ variables. For any $n\ge m$, the function $f^*$ defined by the CA construction of Equation~\eqref{eq:constr} has the same algebraic degree of $f$.
\begin{proof}
The result is clearly true when $n=m$, since in that case $f \equiv f^*$. We thus only consider the case where $n>m$.

Let $d$ be the algebraic degree of $f$. Each summand in Equation~\eqref{eq:constr} has degree $d$, since it always corresponds to the local rule $f$ applied on a different neighborhood. We thus have to show that not all terms of degree $d$ cancel each other out. Consider the first summand $f(x_1,\cdots, x_m)$, and let $S_d = \{I \subseteq 2^{[m]}: |I| = d, a_I \neq 0\}$ be the set of monomials of degree $d$ in the ANF of $f$. Further, denote by $I_{min} \in S_d$ the minimum element of $S_d$ with respect to the lexicographic order, that is, if $I_{min} = \{i_1,\cdots, i_d\}$ and $J = \{j_1,\cdots,j_d\}$ is any other set of $S_d$, it holds $i_k < j_k$ for some $k \in [d]$ and $i_h = j_h$ for all $h \in [k-1]$. This monomial cannot be cancelled by any other monomial in the ANF of the subsequent summands, since by Equation~\eqref{eq:constr} their neighborhoods are shifted by at least one coordinate with respect to that of the first summand. Indeed, if we take the $l$-th summand $f(x_{l},\cdots,x_{l+m-1})$ for $l \in \{2,\cdots,n-m+1\}$, and we denote by $I_{min}^l$ its minimum monomial of degree $d$ in lexicographic order, we have that $I_{min}^l = (i_1+l,\cdots,i_d+l)$, which is distinct from $(i_1,\cdots,i_d) = I_{min}$. Hence, the variables in the monomial $I_{min}^l$ cannot overlap completely those of $I_{min}$, which means that the two terms do not cancel each other out. Similarly, the monomial $I_{min}$ cannot be canceled by any non-minimal monomial of degree $d$ in the $l$-th summand. Hence, the monomial corresponding to $I_{min}$ appears in the ANF of~\eqref{eq:constr}, which proves that the algebraic degree of $f^*$ is also $d$. \qed
\end{proof}
\end{lemma}

\section{Search Algorithm}
\label{sec:algorithm}
Lemma~\ref{lm:alg-deg} gives us a first basic insight on the nature of the functions resulting from our construction. However, the fact that the algebraic degree of the original function is preserved is not sufficient from the cryptographic point of view, since as we saw in Section~\ref{sec:background} there are other properties to take into account, such as balancedness and nonlinearity. Considering that semi-bent functions offer a good trade-off of these criteria, we turn our attention to the following question: what are the semi-bent Boolean functions that give rise to an infinite family of semi-bent functions when used as local rules of our CA-based construction? In other words, we are interested in finding a subset of semi-bent Boolean functions of $m$ variables such that they generate semi-bent functions for any number of variables $n \ge m$ when plugged in Equation~\eqref{eq:constr}. In this section and in the next one, we address this question by adopting an \emph{experimental approach}. More precisely, we devise a combinatorial search algorithm to efficiently explore the search space of local rules, and retain only those semi-bent rules over which our construction yields semi-bent functions up to a specified number of variables. Clearly, we cannot prove that the rules obtained in this way indeed generate infinite families of semi-bent functions. However, this experimental search is useful to isolate at least a subset of candidate rules, to be investigated in future research.

A trivial algorithm to search for semi-bent functions simply consists in enumerating all possible truth tables of $m$-variables functions, which are $2^{2^m}$ in total. However, this brute-force procedure is extremely inefficient: most Boolean functions are not semi-bent, and searching through all of them is feasible only up to $m=5$ variables. We thus designed a combinatorial algorithm to exhaustively enumerate only the Boolean functions having a fixed algebraic degree. In this way, by Lemma~\ref{lm:alg-deg} we know that these functions will all generate Boolean functions of the same degree through our construction. This remark is especially useful when considering the case of \emph{quadratic functions}, i.e. functions of degree $2$. As a matter of fact, quadratic functions are a subclass of \emph{plateaued functions}~\cite{carlet10}, which in turn include semi-bent functions, as mentioned in Section~\ref{sec:background}. Hence, focusing on the intersection of quadratic and semi-bent functions is a reasonable trade-off between obtaining an interesting enough class of functions to investigate with respect to our construction and enumerating it in a limited amount of time.

Our search algorithm is based on the ANF representation. Given a target algebraic degree $d$, the $2^m$-bit vector of the ANF coefficients can be easily constrained to yield only Boolean functions of degree $d$: it suffices to set \emph{at least one} of the coefficients $a_I$ such that $|I|=d$ to $1$, while all coefficients $a_J$ with $|J|>d$ must be set to 0. The other coefficients related to monomials of lower degree can be freely chosen. Then, by using the \emph{M\"{o}bius Transform} recalled in Equation~\eqref{eq:mobius}, one can recover the truth table starting from its ANF coefficients, and  check if the corresponding quadratic function is semi-bent by computing its Walsh spectrum. In this case, we can finally test if our construction generates quadratic semi-bent functions up to a specified number of variables. The pseudocode of our search algorithm is reported in Figure~\ref{fig:alg}.

\begin{figure}[t]
\begin{description}
\item[{\sc Search-ANF}$(m,n,d)$]
\item[Initialization:] For $1 \le k \le d$, build the family $\mathcal{I}_k=\{I \subseteq [d]: |I| = k\}$ of monomials of degree $k$, set all $2^m$ ANF coefficients of $f$ to 0 and initialize $\mathcal{L}$ as the empty list
\item[Outer Loop:] For all subsets $\mathcal{T} \subseteq \mathcal{I}_d$ (except the empty set), do:
  \begin{description}
    \item[ANF Initialization:] Reset all $d$-degree terms in the ANF to 0
    \item[Instantiation:] For all $T \in \mathcal{T}$, set the ANF coefficient $a_T$ to $1$, i.e. include in the ANF the combination of $d$-degree monomials defined by $\mathcal{T}$
    \item[Inner Loop:] For all subsets $\mathcal{P} \subseteq \bigcup_{k=1}^{d-1} \mathcal{I}_k$ do:
    \begin{enumerate}
        \item Reset all terms of degree less than $d$ in the ANF to 0
        \item For all $P \in \mathcal{P}$, set the ANF coefficient $a_P$ to $1$, i.e. include in the ANF the combination of monomials of degree at most $d-1$ defined by $\mathcal{P}$
        \item Apply the M\"obius Transform (Equation~\eqref{eq:mobius}) to the ANF coefficients vector to obtain the truth table of the function $f$
        \item Compute the Walsh transform (Equation~\eqref{eq:walsh}) on the truth table of $f$
        \item If $f$ is semi-bent, then for all $d < i \le n$ apply the CA construction of Equation~\eqref{eq:constr} with $i$ cells, and compute the Walsh transform of $f^*$
        \item If for all $d < i \le n$ the function $f^*$ is semi-bent, then add $f$ to $\mathcal{L}$
    \end{enumerate}
  \end{description}
\item[Output:] return $\mathcal{L}$
\end{description}
\caption{Pseudocode of the {\sc Search-ANF} algorithm.}
\label{fig:alg}
\end{figure}

\section{Complexity and Search Experiments}
\label{sec:results}
Let us analyze the time complexity of the search algorithm described in the previous section for the case of quadratic functions, i.e. when $d=2$. The outer loop is applied over all subsets of monomials of degree $2$, except the empty set which of course does not give a quadratic function. Since the number of quadratic terms in the ANF of a $m$-variable function is $\binom{m}{2}$, it means that the outer loop is executed $2^{\binom{m}{2}}-1$ times. The inner loop iterates only through all combinations of \emph{linear terms}, hence it is executed for $2^{\binom{m}{1}}$ steps. The search space $\mathcal{S}_{m,2}$ visited by our algorithm is thus composed of the following number of ANF vectors:
\begin{equation}
\label{eq:search-space}
S_{m,2} = \left(2^{\binom{m}{2}}-1\right)\cdot 2^{\binom{m}{1}} = \left(2^{\frac{m(m-1)}{2}}-1\right)\cdot 2^{m} \enspace .
\end{equation}
It follows that $S_{m,2} = 2^{\mathcal{O}(m^2)}$, which is asymptotically better than the $\mathcal{O}(2^{2^m})$ bound given by the brute-force search approach.

We thus applied our algorithm {\sc Search-ANF} on the sets of quadratic functions of $3 \le m \le 6$ variables, testing the CA construction up to $n=20$ variables. Table~\ref{tab:results} reports the results of our search. In particular, for each considered $m$ we give the corresponding number $2^{2^m}$ of $m$-variable Boolean functions which would be searched by a brute-force algorithm, the number $S_{m,2}$ of quadratic functions actually explored by our algorithm and the number $QSB$ of quadratic semi-bent functions found over which our construction works.
\begin{table}[t]
  \caption{Results obtained with the {\sc Search-ANF} algorithm and by filtering only the rules that produce balanced functions.}
  \centering
  \begin{tabular}{p{1cm}p{2cm}p{1.5cm}p{1.5cm}p{1.5cm}}
  \toprule
    $m$ &   $2^{2^m}$                    &   $S_{m,2}$        &   $QSB$     & $Bal$\\ 
  \midrule
    $3$ &   $256$                        &   $56$             &   $24$      & $8$ \\
    $4$ &   $65\,536$                      &   $1\,008$         &   $0$     & $0$ \\
    $5$ &   $\approx 4.3 \cdot 10^9$     &   $32\,736$        &   $2\,208$  & $280$ \\
    $6$ &   $\approx 1.84 \cdot 10^{19}$ &   $2.1 \cdot 10^6$ &   $12\,208$ & $1937$ \\
    \bottomrule
  \end{tabular}
  \label{tab:results}
\end{table}
A first remarkable finding that one can draw from Table~\ref{tab:results} is that our construction does not work on \emph{any} quadratic function of $4$ variables. In particular, the largest number of CA cells for which our construction produced semi-bent functions for $m=4$ variables was $n=8$. Contrarily, for all other values of $m$ our algorithm found semi-bent functions over which our construction worked up to the target value $n=20$. For this reason, we excluded the case $m=4$ in our subsequent experiments.

To further investigate the functions produced by our construction, we considered two additional cryptographic properties: balancedness and number of non-trivial linear structures. Among the functions found by the {\sc Search-ANF} algorithm for which our CA-based construction always produced semi-bent functions of up to $20$ variables, we filtered only those local rules that always produce balanced functions, as reported in the last column of Table~\ref{tab:results}. For each of the remaining functions, we observed that the number of linear structures of every function obtained with the application of our construction is constant. In particular our experiments show that, regardless of the number of variables of the initial local rule, the number of linear structures of each constructed function is equal to $1$ when the number of cells $n$ is odd, and $3$ when $n$ is even.

\section{Conclusions and Open Problems}
\label{sec:conclusions}
As we observed in Section~\ref{sec:algorithm}, our experimental results do not rule out the possibility that our CA-based construction fails for $n>20$ over the semi-bent rules found by our algorithm. However, we believe that at least for a subset of these rules this construction indeed generates semi-bent functions for any $n \in \N$, and the preliminary filtering operation performed in this paper greatly reduces the number of possible candidates, thus easing their theoretical analysis for future research. The first interesting open question to address is understanding why our construction always failed only for $m=4$ variables, and to assess whether this is the case also for other numbers of variables not considered in this work. Then, the next step would be to investigate the rules filtered by our combinatorial search experiments, and try to formally characterize the family of quadratic rules for which our CA-based construction always yields semi-bent functions. A possible idea towards this direction would be to study more in depth the regularity of the number of linear structures of these functions, and assess whether this could be a necessary or sufficient condition for our construction to work.

From an applicative point of view, we remark that the $8$ balanced rules of $m=3$ variables found in our experiments include the elementary rules $30$ and $210$, which have been extensively adopted for designing CA-based cryptographic primitives~\cite{wolfram85,keccak}. It could thus be interesting to investigate whether our construction could enhance these primitives, such as the CA pseudorandom generator in~\cite{wolfram85}, which samples only one cell of a CA with rule $30$ to produce a pseudorandom keystream. Since rule $30$ seems to produce semi-bent functions for any $n \in \N$, one idea could be to modify the pseudorandom generator by taking the value of \emph{all} cells in the CA instead of only the central one, and then compute their XOR as the next pseudorandom bit.

More in general, a very interesting research direction would be to investigate our construction with respect to semi-bent functions of higher algebraic degree. Indeed, even though quadratic functions can reach high levels of nonlinearity, their degree is too low and this can be exploited in \emph{algebraic attacks}~\cite{carlet10}. In this regard, it would be interesting to apply our algorithm to search for cubic semi-bent functions over which our construction works.

Finally, another venue for further research on a topic which is not related to cryptography but rather on the theory of CA themselves, is to investigate whether our construction could give any insight about the periods of \emph{spatially periodic preimages} in surjective CA. As shown in~\cite{mariot17}, the least periods of preimages families in \emph{bipermutive} CA are characterized by disjoint cycles, but up to now an algebraic characterization of such periods has been given only for the case of linear rules. Given that the rules of an even number of variables found in our experiments are all characterized by three linear structures, it could be the case that many of them are indeed bipermutive, considering the connection between bipermutivity and linear structures observed for instance in~\cite{leporati14}. In that case, our construction could possibly give further information on the least periods of preimages of quadratic bipermutive CA.

\section*{Appendix: Source Code and Experimental Data}
The source code of the search algorithm and the experimental data are available at \url{https://github.com/rymoah/ca-boolfun-construction}.

\subsubsection*{Acknowledgements.}
The authors wish to thank Claude Carlet and Stjepan Picek for useful comments on a preliminary version of this work. This research was partially supported by FRA 2020 - UNITS.


\bibliographystyle{abbrv}
\bibliography{references}

\end{document}